\documentclass[twocolumn,showpacs,amsmath,amssymb,aps,superscriptaddress,nofootinbib,longbibliography,notitlepage]{revtex4-1}

\usepackage{graphicx}
\usepackage{xcolor}

\usepackage{soul}
\usepackage{setspace}
\usepackage{enumitem}

\usepackage[colorlinks=true, linkcolor=blue, citecolor=gray]{hyperref}
\usepackage[capitalize]{cleveref}

\usepackage{amsmath}
\usepackage{amssymb}
\usepackage{bbold}

\usepackage{qcircuit}

\usepackage{titlesec}
\titleformat{\paragraph}[runin]{\bf}{\theparagraph.}{1em}{}[.]
\titlespacing*{\paragraph}{2em}{1em}{1em}

\newtheorem{lem}{Lemma}

\newcommand{\bra}[1]{{\langle{#1}\rvert}}
\newcommand{\ket}[1]{{\lvert{#1}\rangle}}
\newcommand{\braket}[2]{{\langle{#1}\rvert{#2}\rangle}}
\newcommand{\ketbra}[1]{\ket{#1}\bra{#1}}
\newcommand{\expval}[1]{{\langle{#1}\rangle}}

\renewcommand{\Im}{\operatorname{Im}}
\renewcommand{\Re}{\operatorname{Re}}

\DeclareMathOperator*{\argmin}{arg\,min}
\DeclareMathOperator{\Var}{Var}
\DeclareMathOperator{\Cov}{Cov}
\DeclareMathOperator{\sgn}{sgn}
\DeclareMathOperator{\Tr}{Tr}

\begin{document}

\title{Optimizing the information extracted by a single qubit measurement}

\begin{abstract}
We consider a quantum computation that only extracts one bit of information per $N$-qubit quantum state preparation.
This is relevant for error mitigation schemes where the remainder of the system is measured to detect errors.
We optimize the estimation of the expectation value of an operator by its linear decomposition into bitwise-measurable terms.
We prove that optimal decompositions must be in terms of reflections with eigenvalues $\pm1$.
We find the optimal reflection decomposition of a fast-forwardable operator, and show a numerical improvement over a simple Pauli decomposition by a factor $N^{0.7}$.
\end{abstract}

\date{\today}

\author{Stefano Polla}
\affiliation{Google Quantum AI, 80636 München, Germany}
\affiliation{Instituut-Lorentz, Universiteit Leiden, 2300 RA Leiden, The Netherlands}

\author{Gian-Luca R. Anselmetti}
\affiliation{Covestro Deutschland AG, Leverkusen 51373, Germany}

\author{Thomas E. O'Brien}
\affiliation{Google Quantum AI, 80636 München, Germany}
\affiliation{Instituut-Lorentz, Universiteit Leiden, 2300 RA Leiden, The Netherlands}

\maketitle

\section{Introduction}

The largest bottleneck in quantum algorithm design is the encoding and decoding of a quantum state.
Although each full characterization of a quantum state requires an exponentially large amount of information, direct measurements of an $N$-qubit quantum state $\rho$ extract only $N$ bits of information, and collapse $\rho$ to a state described by those $N$ bits alone --- erasing any other information. 
Performing this repeatedly allows the estimation of an expectation value $\expval{O} := \Tr[O\rho]$ of any operator $O$ that is diagonal in the measurement basis.
The rate at which such a measurement converges is known as the standard quantum or shot noise limit \cite{braunstein1994} - after $M$ repeated preparations, $\expval{O}$ can be estimated with variance
\begin{equation} \label{eq:VarO}
    \Var[O] = 
    M^{-1}\left(\expval{O^2} - \expval{O}^2\right).
\end{equation}
Though this rate can be improved upon~\cite{Giovannetti2004, Higgins2009, Knill2007, Huggins2021}, doing so requires implementing long coherent circuits or performing large correlated measurements, which are not feasible in the current NISQ era~\cite{Preskill2018}.

Instead of using all $N$ qubits to extract data from a quantum state, one may perform a partial measurement that extracts less than $N$ bits, and use the remaining qubits to detect and mitigate errors~\cite{Bonet-Monroig2018, McArdle2019, Huggins2019}.
Error mitigation is key in obtaining precise results from NISQ circuits, such as variational algorithms~\cite{Peruzzo2014, McClean2016}, where the output of the quantum algorithm is a set of estimates of expectation values. 
Echo verification (EV - see Section~\ref{sec:EV}) \cite{OBrien2021, Cai2021, Huo2022, obrien2022} allows one to strongly mitigate errors in a wide class of algorithms, by recasting measurements as Hadamard tests.
In each EV circuit, a single bit of information is extracted from the system register as a measurement, freeing up the remainder of the register for error detection/mitigation. 
One may combine results of multiple EV circuits (through classical post-processing) into an error-mitigated estimator of any target quantity.
However, the stringent requirement that only one bit of information be extracted from the device further tightens the bottleneck of quantum-classical I/O.

In this paper we study how we can optimize information extraction from a quantum system to estimate the expectation value of an observable $O$, under the restriction that only a single bit of information 
is measured per state preparation.
This matches the requirements of EV, the rest of the information being reserved for error mitigation.
We do not focus in this work on the effectiveness of EV as an error mitigation strategy, and consider only the case of error-free quantum simulation.
We define measurements with a single-bit outcome in terms of the Hadamard test, use these to construct an expectation value estimator for a more complicated operator via a linear decomposition, and calculate the variance of this resulting estimator.
We prove necessary conditions for such a linear decomposition to be optimal; i.e. to minimize the cost of expectation value estimation.
We construct a provably optimal (in some sense) decomposition for a fast-forwardable operator, and give a general (albeit expensive) method to implement this decomposition through quantum signal processing \cite{Low2017, Low2019, Gilyen2019}.
We analyse our methods numerically, comparing the variance of estimators based on our optimal method with other known approaches such as Pauli decompositions and the Dirichelet kernel measurements introduced in \cite{Wierichs2021}.
We find an asymptotic improvement between our optimal decomposition and a simple Pauli decomposition of a factor $N^{0.7}$, which at $13$ qubits gives already an order of magnitude improvement.

\section{Single-qubit measurements}

The most general measurement that extracts one bit of information from a $N$-qubit state $\ket\psi$ is a binary Positive-Operator Valued Measurement (binary POVM); this is defined by two positive operators $\Pi_+, \Pi_- > 0$ such that $\Pi_+ + \Pi_- = \mathbb{1}$.
The outputs of such measurement, which we label $+1$ and $-1$, have probabilities $p_\pm = \bra\psi \Pi_\pm \ket\psi$.
Schematically,
\begin{equation*}
    \includegraphics{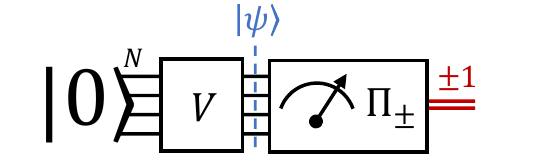},
\end{equation*}
where we defined the unitary preparing the state $V\ket{0}:=\ket{\psi}$.
In section \ref{sec:HT}, we review the Hadamard test and we show that there exists a one-to-one equivalence between outcomes of Hadamard tests and binary POVMs.

Extracting only a single bit allows further processing of the quantum information remaining in the state register.
For instance, inverting the unitary that prepared $|\psi\rangle$ and measuring in the computational basis yields a powerful error mitigation technique, echo verification~\cite{OBrien2021, Cai2021, Huo2022}, which we review in Section~\ref{sec:EV}.
In another example, the Hadamard test may be used to estimate the gradient of a cost function with respect to a variational term $\exp(iA\theta)$ in a circuit, as $\frac{d}{d\theta}\exp(iA\theta)=iA\exp(iA\theta)$~\cite{Guerreschi2017, Li2017}.
Both these methods require operating on the system register after the binary measurement is performed, preventing further information extraction. 
(For the specific case of EV, we show in App.~\ref{app:EV_parallel} that extracting more than one bit of information is counterproductive.)
Furthermore, this restricted output model of quantum computation can be relevant in quantum-enhanced metrology settings \cite{Giovannetti2004, giovannetti2006}, where a single-qubit probe is used \cite{saunders2021}. 
A similar restricted access model has been studied in the context of Hamiltonian learning \cite{di2009,burgarth2011}.
Note that this single qubit access model is different to the one clean qubit model of computation (DQC-1)~\cite{knill1998}; here we consider using a single qubit to extract information from a non-trivial quantum state.

\subsection{The Hadamard test}
\label{sec:HT}

A Hadamard test (HT) is a binary measurement performed on a state $\ket{\psi}_\text{s}$ in the $N$-qubit system register s. 
It is implemented through a control qubit c initialized in the state $\ket{+}_\text{c}$, a controlled unitary $CU$ and a projective Pauli measurement $X_\text{c}$ on the control qubit.
As a quantum circuit this can be written
\begin{equation*}
    \includegraphics{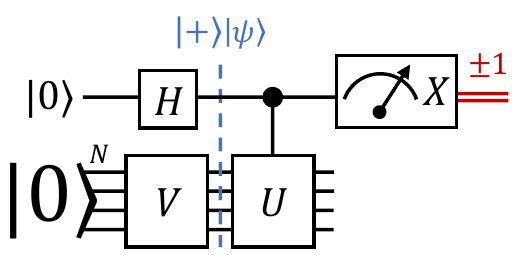}
\end{equation*}
and the resulting state before measurement can be easily calculated to be
\begin{equation}
    \ket\Phi = \frac{1}{\sqrt{2}}\Big(\ket{0}\ket{\psi} + \ket{1} U\ket{\psi}\Big).
\end{equation}
Tracing out the system register then yields the following reduced density matrix on the control qubit,
\begin{equation}
    \rho_\text{c} = \frac{1}{2} \left(\begin{array}{cc}1 & \expval{U} \\ \expval{U}^* & 1\end{array}\right).
\end{equation}
One may estimate the expectation value of $\Re(U):=\tfrac{1}{2}(U+U^{\dag})$ by measuring the control qubit in the $X$ basis, which returns $\Tr[X \rho_\text{c}] = \expval{\tfrac{1}{2}(U+U^{\dag})}$.

To prove the equivalence between HT and binary POVM, we explicitly construct one from another.
To construct the binary POVM corresponding to the HT, we define the measurement operators that represent the back-action of the measurement on the system register
\begin{equation}
    M_\pm = \bra{\pm}_\text{c} CU \ket{+}_\text{c} = \frac{\mathbb{1} \pm U}{2},
\end{equation}
and the relative positive operators $\Pi_\pm = M_\pm^\dag M_\pm$ used to compute probabilities $p_\pm = \bra{\psi} \Pi_\pm \ket{\psi}$ of measuring $\pm 1$ on the ancilla.
Vice versa, given a binary POVM $\{\Pi_+, \Pi_-\}$, we can construct a corresponding Hadamard test by choosing a unitary $U$ that satisfies $\Re(U) = \Pi_+ - \Pi_-$,
\begin{equation}
    U = \exp[i \arccos(\Pi_+ - \Pi_-)]. 
\end{equation}
This is always possible because $\Pi_+ - \Pi_-$ is Hermitian and $ \lVert\Pi_+ - \Pi_-\rVert < \lVert\Pi_+ + \Pi_-\rVert = 1$.
It is easy to check that the Hadamard test constructed from this unitary return the correct positive operators $\Pi_\pm$.

\subsection{Echo verification}
\label{sec:EV}

The name echo verification (EV) refers to a class of powerful error mitigation techniques \cite{OBrien2021, Huo2022, Cai2021}, applicable in most algorithms that make use a Hadamard test to perform measurements on a system register.
This technique was originally introduced by the name of \emph{verified phase estimation}~\cite{OBrien2021} as it considered estimating expectation values of multiple unitaries $U_l = e^{iHt_l}$, with an archetypal application in the context of single-ancilla phase estimation.
However, in this work, we consider the more general expectation-value estimation subroutine yielding $\expval{\Re(U)}$. 
We prefer the name echo verification (used also in~\cite{cai2022,obrien2022}) due to the similarities to a Loschmidt echo.

Echo Verification relies on a key idea: exploiting the information left in the system register after the application of the controlled-unitary operator prescribed by the Hadamard test.
This information is used to detect errors and mitigate their effect on estimated quantities.
This is done by ``echoing'' the preparation unitary $V$, i.e.~applying $V^\dag$ after the controlled evolution, and verifying whether the register s returns to the initial state $\ket{0}$.
The corresponding circuit is
\begin{equation*}
    \includegraphics{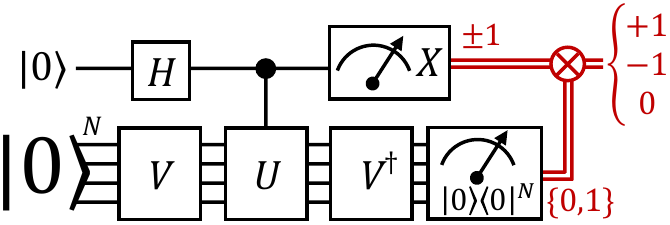},
\end{equation*}
where the multiplication of the classical information channels (red double-lines) sets the circuit output to zero upon failed verification (i.e.~if the final system state is orthogonal to $\ket{0}_s$), and to the output of the Hadamard test otherwise.

Let us denote the combined state after the controlled unitary as $|\Phi\rangle$, and let $\Pi_\psi = \ketbra\psi = V^\dag\ketbra{0}V$ be the projector on the state $\ket\psi_\text{s}$.
The estimate of $\langle\Re(U)\rangle$ can be obtained by measuring the operators $X^{\text{EV}} := X \otimes \Pi_\psi$ on $|\Phi\rangle$ (EV circuit), as opposed to $X_\text{c} := X \otimes \mathbb{1}$ (HT circuit).
One can confirm that, in the absence of error, these operators have identical expectation values on the state at the end of the circuit~\cite{OBrien2021}
\begin{equation}
    \langle\Phi| X^{\text{EV}}|\Phi\rangle = \langle\Phi| X_\text{c} |\Phi\rangle = \langle\psi|\Re(U)|\psi\rangle.
\end{equation}
For an intuitive explanation, note that if the controlled unitary changes the state of the system register, the ancilla qubit must have been in the $|1\rangle$ state, and $\langle 1|X|1\rangle = 0$.
This implies that the expectation value of $X\otimes (\mathbb{1}_N-\Pi_\psi)$ is $0$.

In the presence of a circuit error, verification is likely to fail.
This decreases the expectation value measured by the error probability, which can be measured separately. 
Rescaling the result by the error probability yields a noise-mitigated estimate of the expectation value $\langle\psi|\Re(U)|\psi\rangle$.
The error mitigation power of this method is explored in \cite{OBrien2021, Cai2021, Huo2022, cai2022, gu2022} and experimentally tested in \cite{obrien2022}.
In this work, we only consider noiseless circuits. 

The EV circuit implements a ternary measurement, with outputs $+1, -1, 0$.
Compared to a standard HT defined by the same unitary, the probabilities $p_+$ and $p_-$ are reduced by the same amount ($\frac{p_0}{2}$), yielding a result with the same expected value.
As a consequence, the variance of an EV measurement is always smaller than that of the corresponding HT (this is formalized in Appendix~\ref{app:EV-estimator}).

An extension of Echo Verification allows extracting more than one qubit of information per circuit run by using multiple auxiliary qubits.
However, as the measurement is quadratic in $\ketbra{\psi}$ (resulting by the use of two copies of $V_\psi$ in the circuit), reconstructing the desired expectation values requires nonlinear processing of the measurement results.
Furthermore, as each measurement interferes with the verification of the others, all the variances of estimated expectations increase.
In appendix~\ref{app:EV_parallel} we explore this, and we prove that measuring more than one bit of information per EV experiment is always counterproductive in terms of final variance, for a fixed total number of shots.

\subsection{Ancilla-free echo verification}

The direct (control-based) measurement via the HT may often be replaced by an indirect measurement using an altered circuit~\cite{Mitarai2019, Harrow2021, OBrien2021}, allowing control-free implementations of these single-bit measurements. 
We review briefly the control-free echo verification scheme.

In the Hadamard test, the control qubit provides a clock-reference state $\ket{0}\ket{\psi}$, which is not changed by the application of $CU$.
This clock-reference state is necessary to give physical meaning to the phase $U$ induces on the system register states, thus making it measurable.
If $U$ has a known eigenstate $U\ket{\psi_r} = e^{i\phi_r}\ket{\psi_r}$ orthogonal to $\ket\psi$, this state can be used as a clock-reference removing the need for a control qubit.
In quantum simulation, this state can often be found thanks to the symmetries of the system. 
For example, in second-quantized simulation of particle systems the vacuum state $\ket{0}$ is an eigenstate of any particle-number preserving operator. 

The control-free EV scheme prescribes preparing a cat-state $\frac{1}{\sqrt{2}}(\ket{\psi} + \ket{\psi_r})$, applying $U$, and measuring $X^\text{CFEV} = (\ket{\psi}\bra{\psi_r} + \ket{\psi_r}\bra{\psi})$.
This can be done with the circuit
\begin{equation*}
    \includegraphics{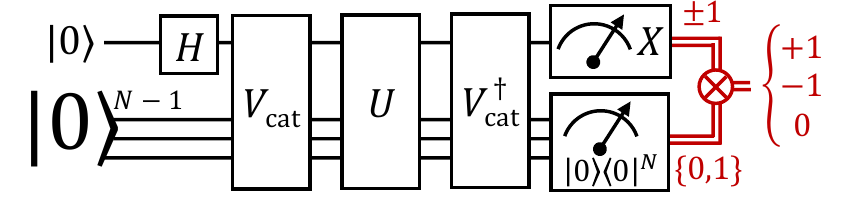}
\end{equation*}
where $V_\text{cat}\ket{0...00} = \ket{\psi_r}$ and $V_\text{cat}\ket{0...01} = \ket{\psi}$.
After the application of $U$, the state is $\ket{\Phi} = \frac{1}{\sqrt{2}}(U\ket{\psi} + e^{i\phi_r}\ket{\psi_r})$, thus
\begin{equation}
    \bra\Phi X^\text{CFEV} \ket\Phi = \bra\psi \Re(U e^{-i\phi_r}) \ket\psi.
\end{equation}
If $\phi_r\neq0$, the desired result $\bra\psi \Re(U) \ket\psi$ can be obtained by substituting $U \to U e^{i\phi_r}$ or applying a phase gate $e^{-i\phi_r/2 Z}$ to the first qubit before measurement.

\subsection{Variance of a binary POVM}

The Hadamard test differs from the projective measurement of $\Re(U):=\tfrac{1}{2}(U+U^{\dag})$ (the Hermitian part of $U$).
Each instance of the Hadamard test can only output $+1$ or $-1$, whereas the spectrum of $\Re(U)$ can have up to $2^{N}$ distinct eigenvalues in the range $[-1, 1]$.
This has a direct impact on the estimation uncertainty: performing the Hadamard test $M$ times and measuring the control qubit in the $X$-basis yields an estimator of $\expval{\Re(U)}=\Re(\expval{U})$ with a variance
\begin{equation} \label{eq:Varstar}
    \Var^*\big[\expval{\Re(U)}\big]
    =
    \frac{1-\expval{\Re(U)}^2}{M},
\end{equation}
which can be seen to be strictly larger than the variance one would obtain by performing a projective measurement of $\Re(U)$ on $M$ copies of $\ket{\psi}$ [Eq.~\eqref{eq:VarO}],
\begin{align} 
    \Var\big[\expval{\Re(U)}\big]
    &\leq
    \Var^*\big[\expval{\Re(U)}\big]\label{eq:VarlessthanVarstar},
\end{align}
as $\expval{\Re(U)^2}\leq 1$.
Our goal is to optimize estimators of expectation values $\expval{O}$ of a given operator, which use data from multiple HTs with different unitaries $U$ [each with the given variance Eq.~\eqref{eq:Varstar}], and assuming one test per state preparation. 
We want to minimize the total number of state preparations (distributed over different choices of $U$) needed to achieve an estimator of $\expval{O}$ with error smaller than a fixed $\epsilon$.

\section{Operator decompositions}

It is common in quantum computing to estimate the expectation value of an operator $O$ by writing $O$ as a linear combination of simpler terms (a.k.a.~a \emph{decomposition}) which have their expectation values estimated independently~\cite{Ortiz2001, Peruzzo2014, Wecker2015}.
In this work, we make use of this method, and consider estimating these simpler terms via Hadamard tests.
Let us fix a decomposition\footnote{In a slight abuse of notation, throughout this work we will use the same label (e.g. $X$) to represent the entire linear decomposition defined by the set $\{c_x, U_x\}$ in Eq.~\ref{eq:Odecomp}, and the set of labels $x$ that we sum over.} $X$,
\begin{equation} \label{eq:Odecomp}
    O = \sum_{x\in X} c_x \Re(U_x) 
    \,\leftrightarrow\,
    \expval{O} = \sum_{x \in X} c_x \expval{\Re(U_x)},
\end{equation}
and consider estimating $\langle O\rangle$ by estimating each $\langle\Re(U_x)\rangle$ independently and summing the results.
As $\Re(U_x)$ and $O$ are Hermitian operators we may assume $c_x$ to be real without loss of generality, and we may further assume $c_x\geq 0$ by absorbing a minus sign onto $U_x$.
Note that the arrow in Eq.~\eqref{eq:Odecomp} points both ways as the set of expectation values on all states $\ket\psi$ uniquely defines an operator.

Once a suitable decomposition $X$ of an operator $O$ [Eq.~\eqref{eq:Odecomp}] has been chosen, to calculate the total cost of the algorithm we must allocate a number $m_x$ of repeated single-shot HT experiments to estimate individual $\langle \Re(U_x)\rangle$.
We assume a single-bit measurement per state preparation, i.e.~each HT requires resetting the circuit and re-preparing $\ket\psi$, and the total number of re-preparations $M_X=\sum_{x \in X} m_x$ is the relevant cost of implementing our measurement scheme.
If each $\expval{\Re(U_x)}$ is estimated independently, the variance on a final estimate of $\expval{O}$ can be calculated by standard propagation of variance
\begin{align}
    \label{eq:propagation_of_variance}
    \Var^*_X\big[\expval{O}\big] 
    &= \sum_{x\in X
    } c_x^2\Var^*\big[\expval{\Re(U_x)}\big]
    \\
    \label{eq:variance_of_decomposition}
    &=\sum_{x\in X}\frac{c_x^2(1-\expval{\Re(U)}^2)}{m_x}.
\end{align}
Eq.~\eqref{eq:VarlessthanVarstar} implies that under the same decomposition of $O$ 
\begin{equation}\label{eq:varObound}
    \Var_X\big[\expval{O}\big]:=\sum_{x\in X}c_x^2\Var[\langle\Re(U_x)\rangle]\leq\Var^*_X\big[\expval{O}\big],
\end{equation}
for all states $\rho$.

\subsection{Adaptive shot allocation} 

Given a decomposition $X$ and a total shot budget $M_X$, an optimal choice for the $m_x$ may be found using Lagrange multiplier methods~\cite{Rubin2018}
\begin{equation} \label{eq:shot-allocation}
    m_x = 
    M_X \frac{
        c_x\sqrt{1-\expval{\Re(U_x)}^2}
    }{
        \sum_{y \in X} c_y\sqrt{1-\expval{\Re(U_y)}^2}
    },
\end{equation}
recalling that $c_x\geq 0$.
This yields a bound on the required $M_X$ to estimate $\expval{O}$ with $\Var^*\big[\expval{O}\big]=\epsilon^2$
\begin{equation} \label{eq:shot_bound}
    M_X \geq \mathcal{M}_X := \epsilon^{-2}\bigg[\sum_{x \in X} c_x\sqrt{1-\expval{\Re(U_x)}^2}\bigg]^2.
\end{equation}
We call $\mathcal{M}_X$ the \emph{cost} of the decomposition $X$.
This may be compared to well-known results for measurement bounds using standard tomography methods~\cite{Wecker2015,Rubin2018,Huggins2019,Verteletskyi2020} by substituting $\Var^*$ for $\Var$ in Eq.~\eqref{eq:propagation_of_variance}.
Though exact values of $\expval{U_x}$ will not be known in advance, these can be estimated using a small initial fraction of measurements before a final distribution of measurements is allocated.

\subsection{The decomposition hierarchy}

We have shown above how to optimize measurement allocation given a linear decomposition $X$ [Eq.~\eqref{eq:Odecomp}].
Let us now consider how to optimize $X$ to minimize Eq.~\eqref{eq:shot_bound}.

We first consider the effect of possible rescalings of $\Re(U_x)$.
If any term $c_x \Re(U_x)$ has $\|\Re(U_x)\|<1$,%
\footnote{Unless stated otherwise, all norms in this work are the spectral norm.}
one can find some unitary $U_{x'}$ for which $\Re(U_{x'})=\Re(U_x)/\|\Re(U_x)\|$; substituting $U_x \rightarrow U_{x'}$ (and $c_x \to c_{x'}$ accordingly) will always improve the bound in Eq.~\eqref{eq:shot_bound}.
(For now we do not worry about how the unitaries may be implemented as quantum circuits; we will consider this issue later.)

One may next consider subdividing individual terms $\Re(U_x)$ of $X$, by writing 
\begin{equation} \label{eq:subdecomposition}
    c_x\Re(U_x)=c_{x,0}\Re(U_{x,0})+c_{x,1}\Re(U_{x,1}),
\end{equation}
where $U_{x,0}$ and $U_{x,1}$ are both unitary, and $c_x, c_{x,0}, c_{x,1}>0$.
As we can assume $\|\Re(U_x)\|=1$, such a decomposition requires $c_{x,0}+c_{x,1}\geq c_x$, to preserve the spectral norm of $\Re(U_{x,0})$ and $\Re(U_{x,1})$.
When this inequality is saturated, we call the sub-decomposition \emph{norm-preserving}.
It turns out that this condition is sufficient for the sub-decomposition to be non-increasing in the cost $\mathcal{M}$ of estimation [Eq.~\eqref{eq:shot_bound}], for all states $|\Psi\rangle$; formally:
\begin{lem}\label{lem:norm_preserving_subdecompositions}
Given a linear decomposition $X$ of a target operator $O$ [Eq.~\eqref{eq:Odecomp}], a sub-decomposition $X'$ [Eq.~\eqref{eq:subdecomposition}] that is norm-preserving has non-increasing cost, $\mathcal{M}_{X'}\leq \mathcal{M}_X$ [Eq.~\eqref{eq:shot_bound}], for any state $|\Psi\rangle$. 
\end{lem}
We give a proof of this lemma in Appendix~\ref{app:lemma1-proof}

We would like to extend the above lemma to a statement that norm-increasing subdecompositions of a linear decomposition $X$ are always suboptimal in some sense.
To achieve this, note that as a corollary to lemma~\ref{lem:norm_preserving_subdecompositions}, we can improve on all terms $c_x\Re(U_x)$ in a linear decomposition $X$ by a norm-preserving identity shift 
\begin{equation} \label{eq:center-subdecomp}
    c_x\Re(U_x) = c_x (1-\bar{\lambda}_x) \Re(U_{\tilde{x}}) + c_x \bar{\lambda}_x \mathbb{1},
\end{equation}
where $\overline{\lambda}_x=\frac{1}{2}(\lambda
^{\min}_x+\lambda^{\max}_x)$, $\lambda^{\min}_x$ and $\lambda^{\max}_x$ are the lowest and highest eigenvalues of $\Re(U_x)$ respectively, and $\Re(U_{\tilde{x}})$ has the same eigenvectors of $\Re(U)$ (with its spectrum shifted and rescaled).
We call the outcome decomposition $\tilde{X}$ of the procedure above the \emph{center} of $X$.
Though a norm-increasing subdecomposition of $X$ may not be suboptimal relative to $X$, it is suboptimal relative to this center:
\begin{lem} \label{lem:norm_increasing_subdecompositions}
Let $X$ be a linear decomposition of $O$ with all $\lVert\Re(U_x)\rVert=1$; let $\tilde{X}$ be the center of $X$ and let $X'$ be a strictly norm-increasing sub-decomposition. 
There exists at least one state $|\Psi\rangle$ for which the cost $\mathcal{M}_{\tilde{X}}<\mathcal{M}_{X'}$.
\end{lem}
We give a proof of this lemma in Appendix~\ref{app:lemma2-proof}.

To recap, the above two lemmas show a) that norm-preserving sub-decompositions do not increase the cost of estimating expectation values via Hadamard tests on any given state, and b) norm-increasing sub-decompositions not only can increase expectation value estimation costs on some states, but are guaranteed to do so on at least one.
This result is in direct contrast to standard expectation value estimation, where independent estimation of $\langle A\rangle$ and $\langle B\rangle$ is sub-optimal to joint estimation of $\langle A+B\rangle$ whenever the latter is possible.
This suggests a path towards optimizing HT expectation value estimation, by repeatedly dividing terms $\Re(U_x)$ in a norm-preserving manner, until no further sub-decomposition can reduce the cost any state.
It turns out that not all choices of division lead to the same end-point, however all end points of this procedure have one common property (proven in Appendix ~\ref{app:lemma3-proof}):

\begin{lem}\label{lem:reflections_are_endpoints}
A decomposition $X$ of an operator $O$ has no non-trivial norm-preserving sub-decompositions if and only if all operators $\mathrm{Re}(U_x)$ in $X$ are reflections: $\mathrm{Re}(U_x)^2=1$.
\end{lem}

It should be no surprise that we find reflection operators $\mathrm{Re}(U_x)^2=1$ to be a crucial ingredient to optimize HT tomography, as these are the only operators that saturate the bound in Eq.~\eqref{eq:VarlessthanVarstar} for all states $|\Psi\rangle$.
We call a decomposition $X$ that consists of reflection operators only a \emph{reflection decomposition}.
We give some simple examples of these in Appendix~\ref{app:example-decompositions}.

\subsection{Optimizing reflection decompositions}

Above we demonstrated that, for a decomposition $X$ of an operator $O$ to be optimal with regards to the cost $\mathcal{M}_X$ of estimating expectation values on a set of states (Eq.~\ref{eq:shot_bound}), all terms in $X$ must be reflection operators.
Otherwise, we demonstrated a means of sub-dividing single terms in the distribution to generate a new distribution with lower cost.
However, this is not to say that all reflection decompositions $X$ have the same cost $\mathcal{M}_X$.
(These two statements are consistent as we cannot transform between reflection decompositions using subdivision.)
The set of reflection decompositions of $O$ form a convex set that is $2^{2^N-N}$-dimensional if all $U_x$ are diagonal in the eigenbasis of $O$.
This raises two questions: is there an optimal decomposition amongst the set of reflection decompositions, and does it achieve the von Neumann bound [Eq.~\eqref{eq:varObound}]?

\begin{lem}\label{lem:Xi_is_optimal}
Let $O$ be an operator and $\Pi_j$ be projectors onto the eigenvalues of $O$; $O\Pi_j=\Pi_jO=\lambda_j\Pi_j$. The \emph{$\Xi$-decomposition} of $O$, given by
\begin{align} \label{eq:xi-decomp}
    O =
    \frac{\lambda_0 + \lambda_J}{2} \, \mathbb{1} 
    \,\,+&\, 
    \sum_{x=1}^{J-1} \frac{\delta \lambda_x}{2} \, \Xi_x
    \\
    \label{eq:xi-elements}
    \Xi_x = \mathbb{1} - \sum_{j<x} 2 \Pi_j
    \,,\quad &
    \delta \lambda_x = \lambda_x - \lambda_{x-1},
\end{align}
uniquely achieves the bound $\mathrm{Var}^*_{\Xi}[O]=\mathrm{Var}[O]$ on all states $|\Psi\rangle$ with support on up to two eigenstates of $O$.
No such decomposition achieves this bound on all states $|\Psi\rangle$ with support on three or more eigenstates of $O$.
\end{lem}
We prove this lemma in Appendix~\ref{app:lemma4_proof}.
Note that the $\Xi$-decomposition can be immediately restricted to any subspace of the full-$2^N$-dimensional Hilbert space containing $|\Psi\rangle$ (i.e. if we knew that due to a symmetry or by virtue of being a low-energy state, $|\Psi\rangle$ had support only on such a space), and the optimality result still holds.
This implies in turn that no linear decomposition $X$ can achieve the von Neumann variance bound even for as small as a $3$-dimensional subspace.
This makes sense, as our restriction to measure one bit of information per state preparation forms a bottleneck with respect to the $3$ nonzero-probability outcomes of a Von Neumann measurement on this space.

\subsection{Implementing the optimal decomposition}

In order to realize the $\Xi$-decomposition estimator, we need to implement HT circuits that (approximately) estimate $\expval{\Xi_x}$.
This may be achieved by realising that
\begin{equation} \label{eq:xi-as-sign}
    \Xi_x = \sgn[ O - \mu_x ]
    \,,\quad
    \mu_x = \frac{\lambda_{x-1} + \lambda_{x}}{2},
\end{equation}
where $\sgn$ is the sign function.
An approximation of this unitary operator can then be realized using quantum signal processing (QSP)~\cite{Low2017,Low2019,Gilyen2019} of the sign function~\cite{Tong2020}, requiring only one additional ancillary qubit.
The QSP circuit is given by
\begin{equation*}
     \Qcircuit @C=1.2em @R=0.6em {
         &\mbox{\hspace{7.5em} repeat for $r=0,...,R-1$}
         \\
         \lstick{\ket{0}_\text{c}} & \gate{R_X(\phi_r)} &
         \multigate{1}{e^{-iZ\otimes(O-\mu_x)t}} & \gate{R_X(\phi_R)} & \qw
         \\
         \lstick{\ket{\psi}_\text{S}} & {/}\qw &
         \ghost{e^{-iZ\otimes(O-\mu_x)t}} & \qw & \qw \mbox{\qquad,}
         \gategroup{2}{2}{3}{3}{1em}{(}
         \gategroup{2}{2}{3}{3}{1em}{)}
    }
\end{equation*} 
where $R_X(\phi_r) = e^{-i \frac{X}{2} \phi_r}$, implements a unitary block encoding $Q_{\boldsymbol{\phi}}$ of a degree-$R$ trigonometric polynomial $S_{\boldsymbol{\phi}}$ of the operator $(O - \mu_x) t$:
\begin{equation}
    \bra{1}_\text{c} Q_{\boldsymbol{\phi}} \ket{0}_\text{c} 
    = \sum_{r=0}^R c_r(\boldsymbol{\phi}) e^{-i r (O - \mu_x)t}
    := S_{\boldsymbol{\phi}}[(O - \mu_x)t].
\end{equation}
Here, $\boldsymbol{\phi}$ is a vector containing the individual angles $\phi_r$ implemented during the QSP circuit.
We can then sample $\expval{\Re\{S_{\boldsymbol{\phi}}[(O - \mu_x)t]\}}$ through HT (or EV), using another qubit controlling all gates in the QSP circuit. 
To approximate Eq.~\eqref{eq:xi-as-sign} with our block-encoded operator $S_{\boldsymbol{\phi}}$, we must choose $t < \frac{\pi}{\lVert O - \mu_x \rVert}$ to avoid aliasing, and find the optimal $\boldsymbol{\phi}$
\begin{equation}
    \boldsymbol{\phi} = \argmin_{(\phi_r = -\phi_{R-r})}
    \int_{0 + \delta}^{\pi - \delta} d\omega \Big[
        \sgn(\omega) - \Im[S_{\phi}(\omega)] 
    \Big].
\end{equation}
Here, the constraint $\phi_r = -\phi_{R-r}$ ensures
$\Im[S_{\phi}(\omega)]$ is an odd function of $\omega$. A resolution parameter $\delta \geq 0$ can be introduced to improve the approximation away from the nodes $\omega=\{0, \pm\pi\}$ of $S_\phi(\omega)$.
In Appendix \ref{app:sign-decomposition} we give further details of this decomposition, and analyse the approximation error numerically.
We find that this error converges exponentially in the number of circuit blocks $R$.

\section{Numerical experiments}

\begin{figure}[t]
    \includegraphics[width=1\columnwidth]{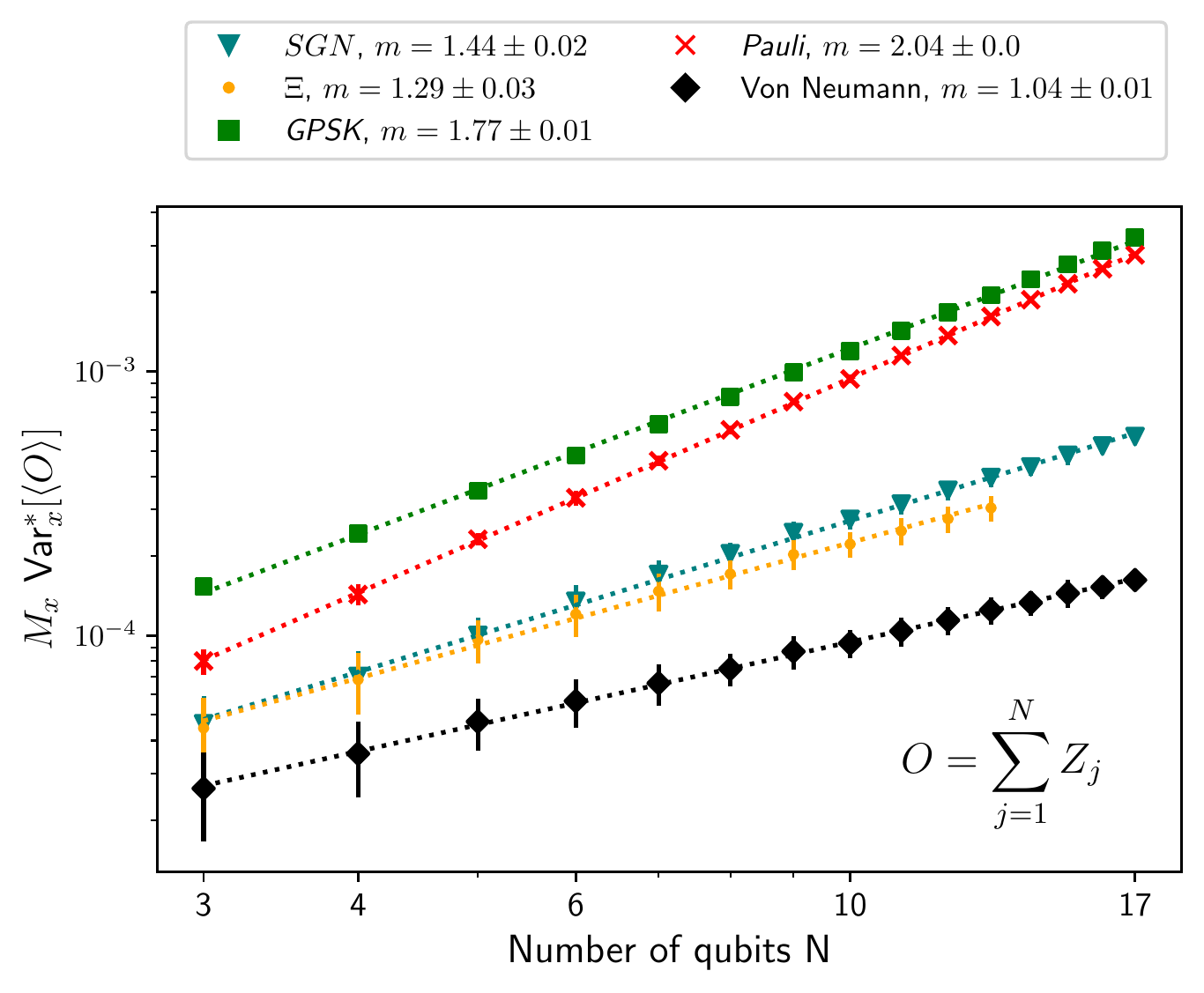}
    \caption{Comparison study of variances of different decompositions on random states generated by a hardware-efficient ansatz (see text for details). Different colours correspond to different decompositions [Eq.\eqref{eq:Odecomp}] of the target operator $O$ (see text for the description of all decompositions). Dashed lines are power-law fits to the data (obtained exponents are given in legend).}
    \label{fig:variances-constant}
\end{figure}

To investigate performance of various decompositions on states that have support on more than two eigenstates of $O$, and therefore are not covered by Lemma \ref{lem:Xi_is_optimal}, we perform numerical simulations using random variationally-generated states and a simple toy operator $O=\sum_jZ_j$.
(In appendix~\ref{app:numerics}, we report this scaling for other systems.)
We measure the variances on states generated by a hardware-efficient ansatz \cite{kandala2017} with random input parameters using PennyLane \cite{pennylane}. 
For each datapoint $100$ random states are generated.
We consider estimating $\expval{O}$ in a realistic scenario where the $\expval{\mathrm{Re}(U_x)}$ values will not be known in advance to optimally choose $m_x$ via Eq.~\eqref{eq:shot-allocation}.
Instead, for each random state we generate a prior estimate of each $\expval{\mathrm{Re}(U_x)}$ from $10^5$ measurements of the state, and use these to determine $m_x$ (which are then only approximately optimal).
This leaves the total shot count $M_X$ as a free parameter; we resolve this in Fig.~\ref{fig:variances-constant} by calculating $M_X\mathrm{Var}^*_X[\expval{O}]$.
(This gives a quantity that is relevant regardless of the number of the shots actually used to estimate $\expval{O}$.)

An average of $M_X\mathrm{Var}^*_X[\expval{O}]$ over the $100$ states is formed and plotted in Fig.~\ref{fig:variances-constant} for each grouping method.
This is compared to the Von Neumann measurement variance $\Var[O]$, which does not require any shot allocation, and sets a lower limit to the other estimators [see App.~\ref{app:lemma4_proof}, Eq.~\eqref{eq:var_bound_decomp}]. 
The $\Xi$-decomposition [orange, `$\Xi$'] has the best asymptotic scaling of all decompositions, being suboptimal to $\Var[O]$ by a factor $\approx N^{1/3}$.
The QSP approximation of $\Xi$, [teal, `SGN'], has a slightly worse asymptotic scaling, which we associate to the error in approximating $\sgn(O-\mu_j)$.
At the largest considered $N=13$, these two decompositions suffer approximately a factor $2$ penalty in their total cost compared to $\Var[\expval{O}]$.
The generalized parameter-shift kernel decomposition \cite{Wierichs2021} [green, `GPSK', described in Appendix \ref{app:GPSK}] has the worst overall performance out of the investigated estimators, due to the constant factor.
It has however a better asymptotic scaling than a simple Pauli decomposition $U_x=Z_j$ [red, `Pauli', Appendix \ref{app:example-decompositions}].
In Appendix~\ref{app:numerics} we investigate the scaling of different sets of observables.
We observe that the order of the performance of the different decompositions remains consistent throughout, but the relative gains and losses in performance can be significantly different.

\section{Conclusion}

In this work we studied the optimization of expectation value estimation for a quantum state in the case where we are only allowed to measure a single qubit per state preparation (e.g. through Hadamard tests, with relevant application to echo verification).
We calculated the cost of estimating the expectation value of an operator $O$ by linearly decomposing $O$ into a linear combination of sub-unitary terms, assuming an optimal shot allocation.
We demonstrated that this cost is strictly non-increasing when terms are further subdivided, under the constraint that this subdivision preserves the induced 1-norm of the term coefficients.
We showed that the end-points of this procedure of repeated division are linear decompositions of $O$ where all terms are reflection operators; a so-called `reflection decomposition'.
We identified one such decomposition, the $\Xi$-decomposition, as unique in its ability to estimate $\langle O\rangle$ with a variance matching the Von Neumann measurement limit on any linear combination of up to $2$ eigenstates of $O$.
We demonstrated how the $\Xi$-decomposition may be approximately implemented through quantum signal processing.
Numerical results demonstrate that on simple systems, the $\Xi$-decomposition and its approximate couterpart demonstrate clear constant and asymptotic improvements over other reflection decompositions (in the cost of estimating $\langle O\rangle$ on random states), with up to a factor $10\times$ improvement for estimation on $20$ qubits.

Though these results are encouraging, the significant discrepancy between $\mathrm{Var}^*_{\Xi}[O]$ and $\mathrm{Var}[O]$ is worrying for NISQ algorithms that already incur a significant cost to tomograph complex Hamiltonians~\cite{Wecker2015, Izmaylov2018, Huggins2019, Bonet-Monroig2018, Crawford2021, Yen2022}; either one incurs a large overhead for measurement due to the need to invoke quantum signal processing or incur the clear asymptotic scaling cost that comes with measuring single Pauli terms per state preparation.
Given that echo verification has a sampling cost scaling as $1/F^2$ (for a circuit fidelity $F$)~\cite{OBrien2021}, this result adds to the unlikelihood of beyond-classical NISQ variational algorithms in chemistry.
Finding reflection decompositions with lower circuit depth is a clear avenue for future work.

\acknowledgments{The authors wish to acknowledge Fotios Gkritsis for providing shot allocation code, and William Huggins, Ryan Babbush, Jordi Tura, Alicja Dutkiewicz, David Wierichs and Christian Gogolin for useful discussions and advice on this work. Covestro acknowledges funding from the
German Ministry for Education and Research (BMBF)
under the funding program quantum technologies as part
of project HFAK (13N15630).}

\bibliography{bibliography.bib}
%

\appendix

\section{Echo verification estimators}
\label{app:EV-estimator}
The estimator used for echo verification is not identical to the one studied in the main text, and so its variance is not quite identical.
In particular, we have $(X^{\text{EV}})^2 = \Pi_\psi$
which implies that the variance on an estimate of $\langle\Phi|X_{\text{EV}}|\Phi\rangle$ is
\begin{align} 
     \Var^*_{\mathrm{EV}}[\expval{\Re(U)}] = \frac{\bra\Phi \mathbb{1} \otimes \Pi_\psi \ket\Phi - \bra\psi \mathrm{Re}(U)\ket\psi^2}{M}.
\end{align}
Clearly $|\braket{\Phi}{\psi}|^2 \leq 1$, which implies $\mathrm{Var}_{\mathrm{EV}}^*[\mathrm{Re}(U)]\leq \mathrm{Var}^*[\mathrm{Re}(U)]$ [by comparison with Eq~\eqref{eq:Varstar}]. In other words, the varianc of the EV estimator is always smaller or equal to the variance of the relative HT estimator.
It is easy to calculate from the circuit above that
\begin{align}
    \bra\Phi \Pi_\psi \otimes \mathbb{1} \ket\Phi = \frac{1}{2}|1 + \bra\psi U\ket\psi|^2,
\end{align}
(noting that $\expval{\Re(U)} = \Re(\expval{U})$,
which can be subtituted back into our variance estimate to obtain
\begin{align}
    \mathrm{Var}^*_{\mathrm{EV}}[\mathrm{Re}(U)]
    &\geq\frac{1-\langle\psi|\mathrm{Re}(U)|\psi\rangle^2}{2M},
\end{align}
Thus, we have 
\begin{equation}
\mathrm{Var}^*[\langle \mathrm{Re}(U)\rangle]\geq \mathrm{Var}_{\mathrm{EV}}^*[\langle \mathrm{Re}(U)\rangle]\geq\frac{\mathrm{Var}^*[\langle \mathrm{Re}(U)\rangle]}{2}.
\end{equation}
This justifies our focus in the main text on optimizing the estimator from a standard Hadamard test; this estimator is simpler to analyse, more general, and differs from the EV estimator (that motivated this work) by at most a factor $2$.

\section{Parallelizing echo verification}
\label{app:EV_parallel}

In absence of echo verification, we can trivially parallelize Hadamard tests measuring $K$ commuting operators $\{\Re(U_0),...,\Re(U_{K-1})\}$ using $K$ ancillary qubits, one controlling each $U_k$.
If each $U_k$ is controlled by a separate ancillary qubit (labeled $k$, where $C_kU_k$ represents the $k$-th unitary controlled by the $k$-th control qubit), the combined state of the system register s and ancillary qubits after all the unitaries are applied will be
\begin{equation}
    \bigotimes_k C_kU_k \ket{+}_k \ket\psi.
\end{equation}
The probabilities of obtaining $\pm 1$ when measuring $X$ on the $j$-th control qubit are
\begin{align}
    p_{j\pm} &= \left\lVert \bigotimes_{k\neq j} C_kU_k \ket{+}_k \frac{1 \pm U_j}{2} \ket\psi \right\rVert^2 \\
    &= \frac{1}{4} \bra\psi (1 \pm U_j^\dag)(1 \pm U_j)\ket\psi
\end{align}
which coincides with the probabilities of a single Hadamard test with unitary $U_j$.

When performing echo verification, parallelization is more complicated.
The result of verification (the measurement of $\Pi_\psi = \ketbra{\psi}$ on the system register) is affected by all the controlled-$U_k$, and thus its result cannot be simply associated to one specific ancilla being in the state $\ket{1}$.
To mitigate errors, all the cases in which the register is found in a state orthogonal to $\ket\psi$ should be considered as null towards all of the ancilla measurement results.
The echo-verified probability of measuring the binary string $\vec{\sigma} = (\sigma_0,...,\sigma_k)$, where each $\sigma_k$ is $\pm1$ corresponding to the state $\ket\pm$ measured on the $k$-th ancilla, is then
\begin{align} \label{eq:parallel-ev-probabilities}
    p_{\vec{\sigma}}^\text{EV} &= \left\lvert \bra\psi \prod_{k} \bra{\sigma_k}_k C_kU_k \ket{+}_{k} \ket\psi \right\rvert^2
    \nonumber \\ &=
    \frac{1}{4^K} \left\lvert \bra\psi \prod_{k} (1 + \sigma_k U_k) \ket\psi  \right\rvert^2.
\end{align}
The product in this equation can be then developed into a linear combination of $2^K$ expectation values (note that, as all $U_k$ commute, the order does not matter). Under the assumption that all these expectation values are real [granted if $U_k = \Re(U_k)$] Eq.~\eqref{eq:parallel-ev-probabilities} defines a quadratic system of $2^K$ equations with $2^K-1$ unknowns\footnote{In the case of a more general $U = \Re(U) + i \Im(U)$, a similar system can be constructed by measuring each $U_k$ and $i U_k$ with $2K$ ancillas. Showing this is besides the scope of our work, and for the sake of simplicity we restrict ourselves to the case of Hermitian $U_k=\Re(U_k)$.}.
Solving such system we find that the expectation value of a single $\Re(U_j)$ can be estimated by processing the sampled $p^\text{EV}_{\vec{\sigma}}$ as
\begin{equation}
    \label{eq:parallel-ev-ReU-estimator}
    \expval{\Re(U_j)} = 
    \underbrace{
        \left(\sum_{\vec{\sigma}: \sigma_j = +1} \sqrt{p_{\vec{\sigma}}^\text{EV}}\right)^2 
    }_{p_{j+}^\text{EV}}
    -
    \underbrace{
        \left(\sum_{\vec{\sigma}: \sigma_j = -1} \sqrt{p_{\vec{\sigma}}^\text{EV}}\right)^2
    }_{p_{j-}^\text{EV}},
\end{equation}
where we denoted $p_{j\pm}^\text{EV}$ the terms that reproduce the probabilities that would be returned by a single, un-parallelized EV experiment
\begin{equation}
    p_{j\pm}^\text{EV} = \frac{1}{4}\left| \bra\psi 1 \pm U_j \ket\psi \right|^2.
\end{equation}
We assume $p^\text{EV}_{\vec{\sigma}}$ are sampled by averaging $M$ shots of the parallel EV experiment. 
These are probabilities of mutually-exclusive measurements, thus the covariance matrix of the  $p^\text{EV}_{\vec{\sigma}}$ estimators is defined by
\begin{align}
    &\Var[p^\text{EV}_{\vec{\sigma}}] 
    = \frac{1}{M} p^\text{EV}_{\vec{\sigma}} (1 - p^\text{EV}_{\vec{\sigma}}),
    \\
    &\Cov[p^\text{EV}_{\vec{\sigma}},
    p^\text{EV}_{\vec{\rho}}] 
    = 
    - \frac{1}{M} p^\text{EV}_{\vec{\sigma}} p^\text{EV}_{\vec{\rho}} 
    \quad\text{if}\quad \vec\sigma \neq \vec\rho.
\end{align}
We can then propagate the error through Eq.~\eqref{eq:parallel-ev-ReU-estimator} to obtain the variance on the parallel-EV (PEV) estimator of $\expval{\Re(U_j)}$
\begin{align}
	M&\Var^*_\text{PEV}[\expval{\Re(U_j)}]
	= \nonumber \\
 	&
	\sum_{\vec\sigma} \frac{p_{j\sigma_j}^\text{EV}}{p_{\vec\sigma}^\text{EV}} p^\text{EV}_{\vec\sigma} (1 - p^\text{EV}_{\vec\sigma}) 
	-
	\sum_{\vec\sigma\neq\vec\rho} \sigma_j \rho_j
	\frac{\sqrt{p_{j\sigma_j}^\text{EV}}}{\sqrt{p_{\vec\sigma}^\text{EV}}}
	\frac{\sqrt{p_{j\rho_j}^\text{EV}}}{\sqrt{p_{\vec\rho}^\text{EV}}}
	p^\text{EV}_{\vec\sigma} p^\text{EV}_{\vec\rho} \nonumber
	\\
	= & \nonumber
	\sum_{\vec\sigma} p_{j\sigma_j}^\text{EV} - \expval{\Re(U_j)}
	\\
	= & 
	2^{K-1} (p_{j\sigma_+}^\text{EV} + p_{j-}^\text{EV}) 
	- \expval{\Re(U_j)}.
\end{align}
which explodes exponentially with the size of the parallelization $K$.

More generally, we can compute the covariance matrix for all the $p_{j,\sigma_j}^\text{EV}$ through error propagation
\begin{align}
    \Var[p_{j\sigma_j}^\text{EV}] &= p_{j\sigma_j}^\text{EV} (2^{K-1} - p_{j\sigma_j}^\text{EV})
    \\
    \Cov[p_{j\sigma_j}^\text{EV}, p_{k\rho_k}^\text{EV}] &=
    \delta_{j,k}\sqrt{p_{j\sigma_j}^\text{EV}p_{k\rho_k}^\text{EV}}
    - p_{j\sigma_j}^\text{EV}p_{k\rho_k}^\text{EV}
\end{align}
[where the covariance assumes $(j,\sigma_j)\neq(k, \rho_k)$].
This shows that, increasing $K$, we effectively add to the covariance matrix a positive semi-definite term with a norm that scales exponentially in $K$. As all the decompositions Eq.~\eqref{eq:Odecomp} are ultimately to be estimated as linear combinations of the sampled probabilities $p_{j\sigma_j}^\text{EV}$, parallelizing error verification is counterproductive.

\section{Proof of decomposition optimality hierarchy}

In this section we build up to the proof that the $\Xi$-decomposition is optimal in terms of cost \eqref{eq:shot_bound}, by proving the lemmas introduced in the main text. 
We first prove that a norm-preserving sub-decomposition has non-increasing cost with respect to its parent decomposition, for all states $\ket\psi$. 
We then prove that a sub-decomposition that does not have the norm-preserving property is always sub-optimal (i.e. it has strictly greater cost than an alternative norm-preserving sub-decomposition).
The iteration of the norm-preserving sub-decomposition procedure leads to one of many alternative improving sequences of decompositions. 
The endpoint of each sequence is a norm-preserving linear decomposition of $O$ for which all unitaries are reflection operators.
Finally, we prove that one of such decompositions (the $\Xi$-decomposition) achieves the Von-Neumann measurement variance bound on a certain set of states, and that no unbiased estimator based on single-qubit measurements can achieve this bound on a larger set of states.

\subsection{Proof of Lemma~\ref{lem:norm_preserving_subdecompositions}, and corollaries}
\label{app:lemma1-proof}

Given a linear decomposition $X$ of an operator $O$ [Eq.~\eqref{eq:Odecomp}], 
consider a norm-preserving sub-decomposition $X'$ where a single term $x \in X$ is split according to Eq~\eqref{eq:subdecomposition}. The bound on the total number of shots Eq.~\eqref{eq:shot_bound} will then change:
\begin{align} \label{eq:shots-subdecomposition}
    \mathcal{M}_X 
    &\rightarrow 
    \mathcal{M}_{X'} 
    =
    \epsilon^{-2}\bigg[\sum_{y\neq x}c_y\sqrt{1-\expval{\Re(U_y)}^2}\\&+c_{x,0}\sqrt{1-\expval{\Re(U_{x,0})}^2}+c_{x,1}\sqrt{1-\expval{\Re(U_{x,1})}^2}\bigg]^2.\nonumber
\end{align}
[with the change with respect to Eq.\eqref{eq:shot_bound} being the second row].
This results in a reduction of the cost, as can be seen by calculating
\begin{align} \label{eq:norm-preserving-cost-comparison}
    &
    c_x^2\big[1-\expval{\Re(U_x)}^2\big]\nonumber
    \\ &= 
    (c_{x,0}+c_{x,1})^2-\big(c_{x,0}\expval{\Re(U_{x,0})}+c_{x,1}\expval{\Re(U_{x,1})} \big)^2\nonumber
    \\ 
    &=
    c_{x,0}^2\big[1-\expval{\Re(U_{x,0})}^2\big]+c_{x,1}^2\big[1-\expval{\Re(U_{x,1})}^2\big]\nonumber
    \\ &\;\;
    +2c_{x,0}c_{x,1}\big(1-\expval{\Re(U_{x,0})}^2\expval{\Re(U_{x,1})}^2\big)\nonumber
    \\
    & \geq 
    c_{x,0}^2\big[1-\expval{\Re(U_{x,0})}^2\big]+c_{x,1}^2\big[1-\expval{\Re(U_{x,1})}^2\big]\nonumber
    \\ &\;\; 
    + 2c_{x,0}c_{x,1}\sqrt{\big[1-\expval{\Re(U_{x,0})}^2\big]\big[1-\expval{\Re(U_{x,1})}^2\big]}\nonumber
    \\ &=
    \bigg[c_{x,0}\sqrt{1-\expval{\Re(U_{x,0})}^2}+c_{x,1}\sqrt{1-\expval{\Re(U_{x,1})}^2}\bigg]^2,
\end{align}
where, in the center inequality we have used the fact that for $0\leq a,b\leq 1$,
\begin{equation} \label{eq:trick}
    1-ab \geq \sqrt{(1-a^2)(1-b^2)}.
\end{equation}

As a corollary and example, we look at identity shifts of a term $x \in X$.
For $\Re(U_x)$ with unit norm, we can assume without loss of generality the largest eigenvalue is $\lambda_\text{max} = 1$, and the smallest is $\lambda_\text{min}$. We can then perform the simple norm-preserving decomposition
\begin{equation} \label{eq:max-id-shift}
    c_x\Re(U_x) = c_x (1-\bar{\lambda}) \Re(U_{x'}) + c_x \bar{\lambda} \mathbb{1}
\end{equation}
with $\bar{\lambda} = \frac{1}{2}(\lambda_\text{min} + \lambda_\text{max})$.
The resulting $\Re(U_{x'})$ has maximum eigenvalue $+1$ and minimum eigenvalue $-1$, thus it does not admit non-trivial identity shift.

A norm-preserving sub-decomposition Eq.~\eqref{eq:subdecomposition} of a term with $|\Re(U_x)|=1$ will only admit terms with $|\Re(U_{x,i})|=1$. 
(This can be checked by taking the expectation value of both sides of Eq.~\eqref{eq:subdecomposition} on the eigenstate on which $|\expval{\Re(U_x)}|=1$.)
By the same reasoning, terms with $\Re(U_x)$ having maximum eigenvalue $+1$ and minimum eigenvalues $-1$ [like those obtained by the identity shifts Eq.~\eqref{eq:max-id-shift}] only admit sub-decompositions whose terms have the same property.

\subsection{Proof of Lemma~\ref{lem:norm_increasing_subdecompositions}}
\label{app:lemma2-proof}

In this appendix we compare the costs of two decompositions derived by an original decomposition $X$: the center $\tilde{X}$ where all terms are transformed according to Eq.~\eqref{eq:center-subdecomp}, and the norm-increasing subdecomposition $X'$ where a term $x \in X$ is changed according to Eq.~\eqref{eq:subdecomposition} assuming $c_{x,0} + c_{x,1} > c_x$. 
Remembering that all coefficients are positive $c_y>0$, the cost of each decomposition Eq.~\eqref{eq:shot_bound} is the square of a sum of positive values; the terms in this sum for $y \neq x$ do not change for $X \to X'$, and have a non-increasing value for $X \to \tilde{X}$.
We thus focus only on the term $x\in X$ and the derived ones, highighted here
\begin{align}
    \mathcal{M}_{X'}
    &=  \epsilon^{-2}\bigg[ 
        \overbrace{\sum_{j\in\{0,1\}} c_{x,j} \sqrt{1 - \expval{\Re(U_{x, j})}}}^{m'} + ...
    \bigg]^2,
    \\
    \mathcal{M}_{\tilde{X}} 
    &=  \epsilon^{-2}\bigg[ 
        \underbrace{c_x (1-\bar\lambda_x) \sqrt{1-\expval{\Re(U_{\tilde{x}})}^2}}_{\tilde{m}} + ...
    \bigg]^2.
\end{align}
We now prove there exists a state $\ket{\Psi}$ for which $\tilde{m} < m'$, which implies $\mathcal{M}_{\tilde{X}} < \mathcal{M}_{X'}$.

Let $\ket{\psi_+}$ and $\ket{\psi_-}$ be eigenvectors of $\Re(U_{\tilde{x}})$ with eigenvalue $+1$ and $-1$ respectively. We consider three cases:
\begin{enumerate}
    \item $|\bra{\psi_\sigma} \Re(U_{x,j}) \ket{\psi_\sigma}| < 1$ for at least one combination of $\sigma \in \{+, -\}$ and $j \in \{0, 1\}$. 
    In this case, on the state $\ket{\Psi} = \ket{\psi_\sigma}$ we get $\tilde{m} = 0 < m' \neq 0$.
    \item $\bra{\psi_\sigma} \Re(U_{x,j}) \ket{\psi_\sigma} = \sigma$ for all combinations of $\sigma \in \{+, -\}$ and $j \in \{0, 1\}$. By combining Eq.~\eqref{eq:subdecomposition} and  Eq.~\eqref{eq:center-subdecomp} and taking the expectation value on $\ket{\psi_\sigma}$ we obtain
    $\sigma [c_{x,0} + c_{x,1} - c_x (1-\bar\lambda_x)]= c_x \bar\lambda_x,$
    which implies $c_{x,0} + c_{x,1} = c_x$, violating one of the hypotheses of the lemma.
    \item $\bra{\psi_\sigma} \Re(U_{x,j}) \ket{\psi_\sigma} = (-1)^j \sigma$  for all combinations of $\sigma \in \{+, -\}$ and $j \in \{0, 1\}$.
    We define the state $\ket\Psi = \frac{\ket{\psi_+} + \ket{\psi_-}}{\sqrt{2}}$, on which $\expval{\Re(\tilde{U}_x)} = \expval{\Re(\tilde{U}_{x,0})} = \expval{\Re(\tilde{U}_{x,1})} = 0$. 
    On this state, the costs are $\mathcal{M}_{\tilde{X}} = \epsilon^2 c_x^2 (1-\bar\lambda_x)^2$ and $\mathcal{M}_{X'} = \epsilon^2 (c_{x,0} + c_{x,1})^2$.
    As $\bar\lambda_x \geq 0$ and $c_{x,0} + c_{x,1} > c_x$, $\mathcal{M}_{X'} < \mathcal{M}_{\tilde{X}}$.
\end{enumerate}

\subsection{Proof of Lemma~\ref{lem:reflections_are_endpoints}}
\label{app:lemma3-proof}

In this appendix, we prove that the end-point of norm-preserving decomposition sequences are reflection operators. In other terms, if $\Re(U_x)$ is a reflection operator, it only admits a norm-preserving sub-decomposition [Eq.~\eqref{eq:subdecomposition}] if $\Re(U_{x, 0}) = \Re(U_{x, 1}) = \Re(U_{x})$.

To prove this, consider a state $|\psi\rangle$ in the $+1$ eigenspace of $\Re(U_x)$.
For a norm-preserving decomposition, we must have
\begin{align}
    &c_{x,0} + c_{x,1} = c_x = c_x \langle\psi|\Re(U_{x})|\psi\rangle \nonumber\\
    &= c_{x,0} \langle\psi|\Re(U_{x, 0})|\psi\rangle + c_{x,1} \langle\psi|\Re(U_{x, 1})|\psi\rangle.
\end{align}
As $\|\Re(U_{x,0})\|,\|\Re(U_{x,1})\|\leq 1$, this equality can only be satisfied if $|\psi\rangle$ is also a $+1$ eigenstate of both $U_{x,0}$ and $U_{x,1}$.
A similar argument holds for all $-1$ eigenstates of $U_x$, and so $U_{x,0}, U_{x,1}$ and $U_x$ share the same eigenstates and eigenvalues and must be equal.
Taking such a sub-decomposition has no effect on the estimator of $\expval{O}$, as the same HT are performed and the total number of shots doesn't change, i.e.~$\mathcal{M}_{X'} = \mathcal{M}$ in Eq.~\eqref{eq:shots-subdecomposition}.

\subsection{Examples of reflection decompositions}
\label{app:example-decompositions}

The simplest example of a reflection-based decomposition is a decomposition in terms of Pauli operators
\begin{equation} \label{eq:independent-Z-decomposition}
    O = \sum_j^J c_j Z_j,
\end{equation}
with $c_j \geq 0$.
We could be tempted to measure $\expval{O}$ with a single HT circuit (assuming access to a block-encoding of $\frac{O}{\lVert O \rVert}$, which is optimal). In this case, as $O = \lVert O \rVert \Re(U)$, the bound Eq.~\eqref{eq:shot_bound} is 
\begin{equation} \label{eq:bound-Z-decomp-direct-measurement}
    M \geq \epsilon^{-2} \lVert O \rVert^2 
    \left[ 
        1 - \frac{\expval{O}^2}{\lVert O \rVert^2}
    \right].
\end{equation}
To improve on this, we can estimate each $\expval{Z_j}$ separately, each with a Hadamard test with $U_j = \expval{Z_j}$ (a binary operator). As the spectral norm of $O$ is equal to the induced 1-norm $\lVert O \rVert_1 = \sum_j^J c_j$, Eq.~\eqref{eq:independent-Z-decomposition} is a norm-preserving decomposition.
The bound Eq.~\eqref{eq:shot_bound} then becomes 
\begin{equation}
    M \geq \epsilon^{-2} \left[\sum_j c_j \sqrt{1 - \expval{Z_j}^2}\right]^2,
\end{equation}
which is always smaller or equal than Eq.~\eqref{eq:bound-Z-decomp-direct-measurement} [easily proven through Eq.~\eqref{eq:trick}].
This inequality is only saturated when the considered state $\rho$ has support only on the $\lVert O \rVert^2$-eigenvalue subspace of $O^2$; the operator $O$ projected on this subspace is effectively a binary operator.

Norm-preserving decompositions do not need to involve only mutually commuting Pauli operators.
As a practical example, we consider the two-qubit operator $O = XX + YY$, which appears commonly in quantum Hamiltonians.
As $O = 2 \Im[\mathrm{iSWAP}]$, this operator can be measured with a single Hadamard test circuit. 
Furthermore, in the context of electronic structure Hamiltonians, $O$ preserves particle number, so in general a control-free scheme using the vacuum as reference state can be employed for the measurement.
This operator has three eigenvalues $\{0, \pm1\}$, which means we can improve its measurement by decomposing it in binary operators. 
We propose three decompositions $O = \Re{U_0} + \Re{U_1}$
The obvious Pauli decomposition $U_0 = XX, U_1 = YY$ has the downside of not conserving particle number.
To fix this, we can take 
\begin{equation}
    U_j = \frac{1}{2}[(XX + YY)+(-1)^j(Z\mathbb{1} + \mathbb{1}Z)].
\end{equation}
These are particle-number preserving, reflection operators and can be easily implemented by combining $i$SWAP with single-qubit $e^{\pm i Z \pi/4}$ rotations on both qubits.
The last decomposition,
\begin{equation} \label{eq:xxyy-xi-decomposition}
    U_j = \frac{1}{2}[(XX + YY)+(-1)^j (ZZ + \mathbb{1}\mathbb{1})],
\end{equation}
uses particle-preserving reflection operators with different eigenvalue multiplicities: unlike Pauli operators, the $\pm1$-eigenvalue subspaces of $U_j$ have unequal dimension $1$ and $3$.
For any state in the $0$-eigenvalue subspace, spanned by $\{\ket{00}, \ket{11}\}$, the estimate variance $\Var^*[\expval{\Re(U_j)}]=0$ for decomposition Eq.~\eqref{eq:xxyy-xi-decomposition}. This is not true for the other two decompositions, which indicates that not all decompositions in binary operators are born equal. We will deal with this in the next section.
Another example of a few-qubit reflection operator that is a sum of non-commuting Pauli operators is the three-spin all-to-all Heisenberg coupling
\begin{equation}
    O = \frac{1}{3} \sum_{l=1}^2 \sum_{m=0}^{l-1} X_m X_l + Y_m Y_l + Z_m Z_l,
\end{equation}
which appears e.g.~in the Kagome-Heisenberg Hamiltonian.

\subsection{Proof of Lemma~\ref{lem:Xi_is_optimal}}\label{app:lemma4_proof}

In this appendix we prove Lemma \ref{lem:Xi_is_optimal}, which formally states the optimality and uniqueness of the $\Xi$-decomposition.
To do this, we first define a variance bound for a class of estimators of $\expval{O}$ on a state $\ket\psi$.
We prove that the bound is achieved on all eigenstates of $O$ if all the sampled operators $\Re(U_x)$ are diagonal in the eigenbasis of $O$.
We then construct the $\Xi$-decomposition, and prove that the related estimator saturates the bound on the set $S_2$ of all states with support on at most two eigenstates of $O$.
Finally, we prove no other decomposition satisfies this requirement (i.e. the $\Xi$-decomposition is unique), and no decomposition satisfies the bound on a superset $S \supset S_2$.

A decomposition $X$ [Eq.~\eqref{eq:Odecomp}] of an operator $O$ is optimal on a state $\ket\psi$ if no other decomposition produces an estimator with lower cost [Eq.~\eqref{eq:shot_bound}] for that state.
Optimality can be defined for a set $S$ of states: $X$ is optimal on $S$ if, for each $\ket\psi \in S$, no decomposition $X'$ has lower cost $M_{X'}<M_X$. (Note that this can be readily generalized to mixed state, without changing any of our next results.)
Lemmas \ref{lem:norm_preserving_subdecompositions}-\ref{lem:reflections_are_endpoints} imply a necessary condition for optimality on the whole Hilbert space: $X$ can only be optimal on all states if it has the form
\begin{align}
    \label{eq:necessary-condition}
    O =& \bar{\lambda}_O \mathbb{1} + \sum_{x \in X} c_x \Re(U_x), \\
    c_x>0, \quad
    |\bar{\lambda}_O| &+ \sum_{x \in X} c_x = \lVert O \rVert, \quad 
    \Re(U_x)^2 = \mathbb{1}
    \nonumber
\end{align}
where $\bar{\lambda}_O$ is the average of the largest and smalles eigenvalues of O.
In other words, $X$ is a norm-preserving decomposition of the center of $O$ where all sampled terms are reflection operators.
This condition is not sufficient: as many non-equivalent instances of such decompositions exist, as exemplified in Appendix \ref{app:example-decompositions}. 

We now construct a bound on the variance of the estimator of $\expval{O}$ based on the decomposition $X$: saturating this bound on all $\ket\psi \in S$ implies optimality of $X$ on $S$. [The cost of the decomposition Eq.~\eqref{eq:shot_bound} is defined as the minimum value of $M$ required to achieve target variance $\epsilon^2$, so minimum variance at fixed $M$ implies minimum cost at fixed $\epsilon$.]
\begin{equation} \label{eq:var_bound_decomp}
    \Var^*_X[\expval{O}] 
    = 
    \frac{1}{M}
    \left[
        \sum_xc_x\sqrt{1-\expval{\Re(U_x)}^2}
    \right]^2
    \geq
    \frac{\Var[O]}{M}.
\end{equation}
This bound is implied by Eq.~\eqref{eq:VarlessthanVarstar} and Eq.~\eqref{eq:variance_of_decomposition}, with the choice of optimal shot allocation Eq.~\eqref{eq:shot-allocation}.
It physical interpretation is rooted in the following observation: a Von Neumann measurement of $O$ is the lowest-variance unbiased estimator of $\expval{O}$ when given access to a single state preparation. 
Thus, given $M$ independent experiments each with a single state preparation, the mean of Von Neumann measurements is the lowest-variance unbiased estimator. 

We first consider the set $S_1$ of all eigenstates of $O$. For any $\ket{\phi} \in S_1$, the value of the bound in Eq.~\eqref{eq:var_bound_decomp} becomes $\Var[O] = 0$. 
The bound is thus saturated only if we choose all reflection operators $\Re(U_x)$ diagonal in any eigenvector basis of $O$, i.e. $[U_x, O] = 0$ and $U_x \ket{\phi} = \pm \ket{\phi}$ for any $\ket{\phi} \in S_1$.
For any decomposition of this form, we can write all $U_x$ in terms of the eigenspace projectors of $O$:
\begin{equation}
    \label{eq:diagonal-decomposition}
    U_x = \sum_{j=0}^{J-1} \xi_{x, j} \Pi_j
    \, , \quad \xi_{x, j} \in \{\pm1\},
\end{equation}
where $\Pi_j$ is the projector on the (eventually degenerate) $\lambda_j$-eigenspace of $O$, $J$ is the number of distinct eigenvalues $\{\lambda_j\}$ of $O$, and without loss of generality we assume $\lambda_j > \lambda_{j-1}$.
The coefficients will then have to satisfy the relation $\lambda_j = \sum_x c_x \xi_{x, j}$.

We define the $\Xi$-decomposition based on Eq.~\eqref{eq:diagonal-decomposition}, by choosing $\xi_{x,j} = -1$ if $j<x$, and $+1$ otherwise.
The resulting decomposition is presented in Lemma~\ref{lem:Xi_is_optimal}, Eq.~\eqref{eq:xi-decomp}.
The operators $\Xi_x$ are reflections by definition, and it is easy to check that the decomposition satisfied the necessary condition Eq.~\eqref{eq:necessary-condition}.
Note that $c_0 = (\lambda_0 + \lambda_j)/2$ defines the optimal identity shift (producing the center of $O$) and the $c_x = (\lambda_x - \lambda_{x-1})/2$ complete the decomposition.

We now prove that the $\Xi$-decomposition is optimal on the set $S_2$ of states with support on two eigenstates of $O$, 
\begin{equation}
    S_2 = \left\{
    \frac{\alpha\ket{\lambda_m} + \beta\ket{\lambda_n}}{\sqrt{\alpha^2 + \beta^2}}: 
    \ket{\lambda_m}, \ket{\lambda_n} \in S_1\right\}.
\end{equation}
On a general state $\ket\psi$ with eigenspace occupations $a_j = \bra\psi \Pi_j \ket\psi$, the estimator based on the $\Xi$-decomposition has variance
\begin{equation}
    \Var^*_\Xi[\expval{O}] 
    =    
    \frac{1}{M}
    \left[
        \sum_j^{J-1} \frac{\delta \lambda_j}{2}
        \sqrt{4(\sum_{i<j} a_i) (\sum_{i\geq j} a_i) }
    \right]^2.
\end{equation}
For a state $\ket\phi \in S_2$, only two occupations are nonzero $a_m, a_n \neq 0$ (we assume w.l.g.~$m<n$), thus the term under square root is reduced to $4 a_m a_n$ if $m < j \leq n$ and $0$ otherwise.
The resulting variance 
\begin{align}
    \Var^*_\Xi[\bra{\phi}{O}\ket{\phi}] 
    &=
    \frac{1}{M}
    \left[\frac{\lambda_n - \lambda_m}{2}
        \sqrt{4 a_m a_n}
    \right]^2 
    \\ &= 
    \frac{1}{M} a_n a_m (\lambda_n - \lambda_m) 
    = 
    \Var[\expval{O}] 
    \nonumber
\end{align}
thus saturating the bound Eq.~\eqref{eq:variance_of_decomposition}.

We now prove that the only optimal decomposition on $S_2$ is the $\Xi$-decomposition (or equivalent up to relabeling and trivial subdecompositions).
First of all, $S_1 \subset S_2$, so the terms of the decomposition need to be of the form of Eq.~\eqref{eq:diagonal-decomposition}.
Consider a family of states $\sqrt{a_m}\ket{\lambda_m} + \sqrt{a_n}\ket{\lambda_n}$ for any $n>m$, with only two nonzero eigenstate occupations $a_m + a_n = 1$.
On such a state,
\begin{align}
    \Var^*_X[{O}] 
    & = 
    \frac{1}{M}
    \left[
        \sum_x c_x \sqrt{1 - [a_m \xi_{x, m} + a_n\xi_{x, n}]^2}
    \right]^2
    \nonumber
    \\& 
    = \frac{a_m a_n}{M}
    \left[
        \sum_x 2 c_x \frac{1-\xi_{x, m}\xi_{x, n}}{2}
    \right]^2
    .
\end{align}
The bound Eq.~\eqref{eq:var_bound_decomp} is then saturated when
\begin{align}
    \left[
        \sum_x 2 c_x  \frac{1-\xi_{x, m}\xi_{x, n}}{2}
    \right]^2
    =
    \lambda_n - \lambda_m,
\end{align}
where we  simplified out the free parameter $\frac{a_m a_n}{M}$.
This  can be rewritten as
\begin{equation}
    \sum_x c_x \xi_{x, n} (\xi_{x, n}-\xi_{x, m}) = \sum_x c_x (\xi_{x, n} - \xi_{x, m})
\end{equation}
using the condition on the decomposition coefficients $\lambda_j = \sum_x c_x \xi_{x, j}$.
This implies that, if $\xi_{x, n}=-1$ then $(\xi_{x, m}-\xi_{x, n})=0$ (recall that $c_x > 0$), 
i.e. $\xi_{x, m} = -1$. 
Thus the only $U_x$ that can appear in this decomposition, are of the same form as the operators in the $\Xi$ decomposition ($\xi_{j, m} = -1, \xi_{j, n} = +1$ for $m < j \leq n$), and thus $X$ is either $\Xi$ or a trivial sub-decomposition of it.

We now show that the $\Xi$-decomposition does not saturate the bound Eq.~\eqref{eq:var_bound_decomp} for a state $\ket\psi$ with three non-zero occupations, $a_m, a_n, a_p \neq 0$ ($m<n<p$). On this state we can write
\begin{align}
    \Var^*_\Xi[\expval{O}] =
    \frac{1}{M}\bigg[
        &(\lambda_n - \lambda_m) \sqrt{a_m (a_n + a_p)} \\
        &+ (\lambda_p - \lambda_n) \sqrt{a_p (a_m + a_n)}
    \bigg]^2.
\end{align}
Subtracting from this $\Var[\expval{O}]$, expanding and then collecting terms we get
\begin{align}
    \Var^*_\Xi[\expval{O}] &- \Var[\expval{O}] = \nonumber \\
    = \,\,&\left[(\lambda_n - \lambda_p)(\lambda_n-\lambda_m)\right]\cdot \\
    \cdot &\left[a_m a_p - \sqrt{a_m a_p (a_m + a_n) (a_p + a_n)}\right] > 0 , \nonumber
\end{align}
as both the terms in square brackets are strictly smaller than zero.
This (along with the uniqueness of $\Xi$ as the optimal estimator on $S_2$) implies that no HT-based estimator can saturate the bound Eq.~\eqref{eq:var_bound_decomp} for arbitrary states.

In fact, the bound can only be saturated on states in $S_2$: on these states the Von Neumann measurement has only two possible outcomes ($\lambda_m$ and $\lambda_n$) with nonzero probability.
The adaptive shot allocation scheme then ensures (for a large enough $M$) that most of the measurements we take ($\Xi_x$ with $m\leq x < n$) reproduce the statistics of the Von Neumann measurement, with the single bit we sample in every experiment always distinguishing between $\lambda_m$ and $\lambda_n$.
On any state $\ket\psi \in \overline{S_2}$, the Von Neumann measurement has three or more outcomes with non-zero probability, and we cannot repoduce its statistics by sampling a single qubit per experiment.
This, along with the uniqueness of $\Xi$, implies that no decomposition can satisfy the sufficient condition for optimality on a superset $S \supset S_2$.
The numerical results presented in this paper quantify the increase in variance with respect to the bound, along with confirming the $\Xi$-decomposition outperforms other decompositions on all states.

\section{Implementation of the $\Xi$ decomposition via quantum signal processing}
\label{app:sign-decomposition}

\emph{Verifiable samping of QSP polynomials} --- To measure the operators in the $\Xi$ decomposition Eq.~\eqref{eq:xi-as-sign}, we implement a Hadamard test (or EV) on trigonometric polynomials of $(H - \mu_x)t$ generated by the quantum signal processing. We tune the QSP coefficients such that the polynomials approximate the sign function in a suitable range.
In this section we display and analyse this technique.

The full circuit we use to achieve this is:
\begin{widetext}
\begin{equation*}
     \Qcircuit @C=1.2em @R=0.6em {
         \lstick{\ket{0}_\text{HT}} & \gate{H} & \qw & \qw & \ctrl{1} &
         \ctrl{1} & \ctrl{1} &
         \ctrl{1} & \gate{H} & \meter
         \\
         \lstick{\ket{0}_\text{QSP}} & \qw & \qw & \qw & \gate{R_X(\phi_r)} &
         \multigate{1}{e^{-iZ\otimes(O-\mu_x)t}} & \gate{R_X(\phi_R)} &
         \gate{R_Y(\pi)} & \qw
         & \measureD{\text{(verify)}}
         \\
         \lstick{\ket{\psi}_\text{S}} & {/}\qw & \qw & \qw & \qw &
         \ghost{e^{-iZ\otimes(O-\mu_x)t}} & \qw & \qw & \qw 
         & \measureD{\text{(verify)}} & \mbox{.}
         \gategroup{1}{4}{3}{6}{1em}{(}
         \gategroup{1}{4}{3}{6}{1em}{)}
         \\
         &&&&\mbox{\hspace{7 em} repeat for $r=0,...,R-1$}
    }
\end{equation*} 
\end{widetext}
The first control qubit (labeled HT) takes care of the Hadamard test. 
The second ancilla (labeled QSP) manages the quantum signal processing subroutine, extended through the sign-controlled evolution $e^{-iZ\otimes(O-\mu_x)t}$ to implements a quantum signal processing (QSP) on the operator $e^{(O-\mu_x)t}$.
We now describe how the measurement scheme works, and how to select the $\phi$ parameters to approximate a measurement of $\sgn[(O-\mu_x)t]$ in the interval $[-\pi, \pi]$.

First, we analyze the QSP routine. 
Let us assume $\ket\psi$ to be an eigenstate of $(O-\mu_x)t$ with eigenvalue $\omega \in (-\pi, \pi)$, and only consider the effect of the controlled gates (removing the HT qubit).
Then, we can reduce the cicuit to an effective single-qubit gate on the QSP qubit, with action
\begin{align} \label{eq:qsp-unitary}
    Q_\phi(\omega) &= 
    e^{-i \frac{Y}{2} \pi} 
    e^{-i \frac{X}{2} \phi_{R}} 
    \left[
        \prod_{r=1}^{R} 
        e^{-i \frac{Z}{2} 2 \omega} 
        e^{-i \frac{X}{2} \phi_{R-r}} 
    \right] 
    \nonumber \\
    &= 
    \begin{pmatrix}
    S_\phi(\omega) & \cdot \\
    \cdot     & \cdot
    \end{pmatrix}
\end{align}
which is a block encoding of $S_\phi(\omega)$, a degree-$R$ trigonometric polynomial of $\omega$.
For the sake of simplicity we inserted the final gate $e^{-i \frac{Y}{2} \frac{\pi}{2}} = -i Y$, shifting the polynomial of interest $S$ from the block $\bra{1}Q\ket{0}$ to $\bra{0}Q\ket{0}$.
We ensure $S_\phi(\omega)$ is real and odd by constraining
\begin{equation} \label{eq:real-odd-constraint}
    \phi_r = - \phi_{R-r}
    \implies
    S(\omega) = -S(-\omega) \in \mathbb{R}.
\end{equation}

Re-introducing the system register, i.e. taking a general $\ket{\psi}_\text{S}$, can be done by linearity taking $Q(\omega) \mapsto Q[(O-\mu_x)t]$ and recovering the circuit above.

The result of the verified Hadamard test (or EV) is obtained by measuring on the output state of the circuit the expectation value of $Z_\text{HT}$ (or $Z_\text{HT} \otimes \ketbra{0}_\text{QSP} \otimes \ketbra{\psi}_\text{S}$). (In the absence of noise these two expectation values are equal. In the presence of noise, an additional measurement at $t=0$ can be taken to mitigate errors. For more details on the technique we refer the reader to the original work on EV \cite{OBrien2021}.)

\emph{Approximating the sign function ---} To approximate the operators Eq.~\eqref{eq:xi-as-sign} that make up the $\Xi$ decomposition, we need to choose the QSP parameter $\boldsymbol{\phi}$ such that $S_{\boldsymbol{\phi}}(\omega)$ in Eq.~\eqref{eq:qsp-unitary} approximates $\sgn[\omega]$.
The polynomial $S_{\boldsymbol{\phi}}(\omega)$ is odd, real, and $2\pi$-periodic -- thus having nodes  $S_{\boldsymbol{\phi}}(0) = S_{\boldsymbol{\phi}(\pm \pi)} = 0$. 
To account for the approximation error in the neighborhood of these nodes, we introduce a resolution parameter $\delta \geq 0$, and request the approximation to be effective only in the $[\delta, \pi - \delta]$ interval. 
Choosing $\delta > 0$ implies accepting a larger error in approximating the sign function close to zero. 
For example, we know the eigenvalues of $(O-\mu_x)t$ closest to zero have absolute value $\frac{\delta \lambda_x}{2}t$, we can use this knowledge to choose $\delta$.

\begin{figure}[h]
\centering
\includegraphics[width=0.95\columnwidth]{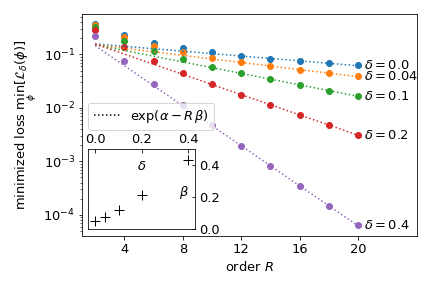}
\caption{
    Loss Eq~\eqref{eq:sign-loss} for the optimal choice of QSP parameters $\boldsymbol{\phi}$, as a function of the order $R$ (number of QSP layers) and resolution parameter $\delta$.
    The dotted lines are log-lin fits for $R>10$. 
    The dependence of the fit parameter $\beta$ on the resolution $\delta$ is shown in the inset. 
}
\label{fig:sign-loss}
\end{figure}

We define a loss function to characterize the quality of the approximation: the average error
\begin{equation} \label{eq:sign-loss}
    \mathcal{L}_\delta(\boldsymbol{\phi}) = \frac{1}{\pi - 2\delta} \int_\delta^{\pi+\delta}
    d\omega \Big[
        \sgn(\omega) - \Im[S_{\boldsymbol{\phi}}(\omega)]
    \Big] .
\end{equation}
To choose the optimal parameters $\boldsymbol{\phi}$, we minimize this loss under the constraints \eqref{eq:real-odd-constraint}.
Although an analytical approach to this problem is possible building on the techniques described in \cite{Gilyen2019}, we take the numerical route to this approximation (which is efficient, scalable and easy to implement).
The integral is thus substituted with a sum on a grid with a number of points much larger than the degree of the trigonometric polynomial.
We plot in Fig~\ref{fig:sign-loss} the minimized cost function, as a function of the approximation's order $R$ and of the resolution parameter $\delta$.
We find that the loss always decays exponentially with an increasing order $R$, with a decay rate depending on $\delta$.

\section{The generalized parameter-shift kernel decomposition of a diagonal operator with ladder spectrum}\label{app:GPSK}

In \cite{Wierichs2021} the authors propose techniques to estimate derivatives $\expval{\frac{d}{dt}U(t)}$ of a unitary $U(t)=e^{iOt}$ generated by $O$, by sampling $\expval{\sin(O t_l)} = \expval{\Re[U(t_l)]}$ at a discrete set of points $\{t_l\}$.
This technique can be used to estimate expectation values of $O$, as $\expval{O} = \big\langle [-i\frac{d}{dt}e^{iOt}]_{t=0}\big\rangle$, and it is clearly compatible with Hadamard test or EV measurements (as it only requires sampling $\expval{\Re{U(t_l)}}$).

For an operator $O$ with equispaced eigenvalues $\Omega, 2\,\Omega, ... , R\,\Omega$ (commonly referred to as a ``ladder spectrum"), the authors give a choice of $\{t_l\}$ and explicit coefficients $c_l(t)$ for the linear combination $\expval{\frac{d}{dt}U(t)} = \sum_l c_l \expval{\Re[-iU(t_l)]}$. 
Assuming $\Omega=1$ (which can be considered a choice of units for the energy), the time points are chosen as $\{t_l = \frac{2 l}{2 R + 1} \pi\}$. 
We can then define a modified version of the Dirichelet kernel,
\begin{equation}
    \tilde{D}_l(t) = 
    \frac{1}{R} \cos(t_l) 
    \left[\frac{1}{2} \sin(R t) + \sum_{j=1}^{R-1} \sin(j t)\right],
\end{equation}
which satisfies $\Tilde{D}_l(t_{l'}) = \delta_{ll'}$. This is a linear combination of the $R$ basis functions $\{\sin(j t)\}_{j=1,...,R}$, like $\expval{\sin(O t)}$. Thus, as the equality
\begin{equation}
    \expval{\sin(O t)} = \sum_{l=1}^R \expval{\sin(O t_l)} \Tilde{D}_l(t)
\end{equation}
holds for all $\{t_l\}_{l=1,...,R}$, it must to hold for all $t$.
We can then differentiate the kernel rather than the expectation value itself. 
Evaluating $[\frac{\partial}{\partial t}\Tilde{D}_l(t)]_{t=0}$ and combining the equations above we obtain
\begin{align}
    \expval{O} 
    &= 
    \sum_{l=1}^R \frac{(-1)^{l-1}}{2R \sin^2(\frac{1}{2}t_l)}  \expval{\sin(O t_l)}
    \\ &=
    \sum_{l=1}^R c_l \expval{\Re[-i U(t_l)]}.
\end{align}
This matches the form of decompositions Eq.~\eqref{eq:Odecomp}. We call thi the generalized parameter shift kernel (GPSK) decomposition.
Under the optimal shot allocation choice [Eq.~\eqref{eq:shot-allocation}], the shot-variance of the estimator based on this decomposition is
\begin{equation}
    M \Var^*_\text{GPSK} = 
    \left[
        \sum_{l=1}^R
            \frac{\sqrt{1 - \expval{\sin(O t_l}^2}}{|2R \sin^2(\frac{1}{2}t_l)|}
    \right]^2
\end{equation}

\section{Details on numerical simulations and further numerical results}
\label{app:numerics}
\begin{figure}[h]
    \centering
    \includegraphics[width=0.95\columnwidth]{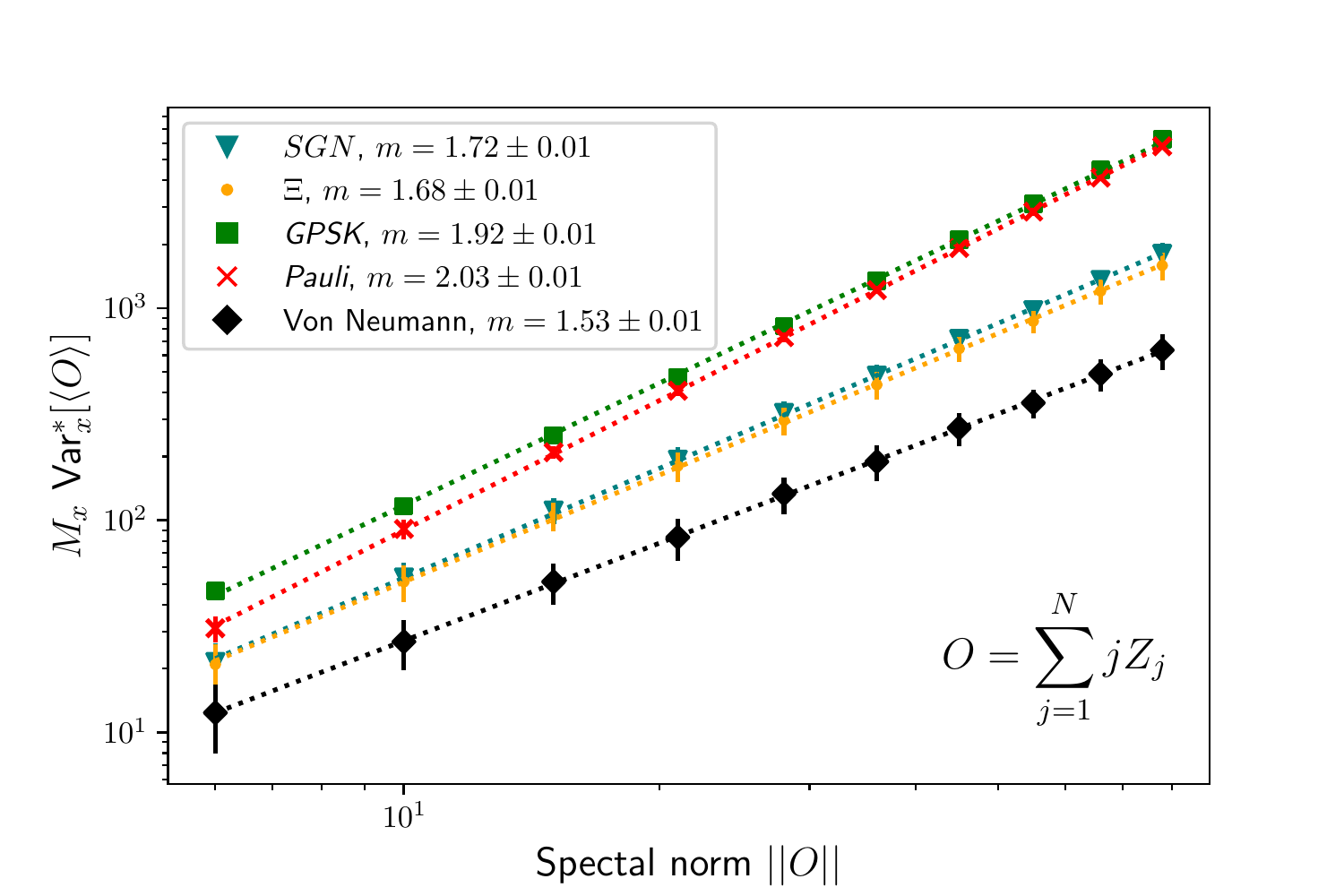}
    \caption{Comparison study of variances of different decompositions on random states generated by a hardware-efficient ansatz (see text for details). Different colours correspond to different decompositions [Eq.\eqref{eq:Odecomp}] of the target operator $O$ (see text for the description of all decompositions). Dashed lines are exponential fits ($a\exp(m N + b)$) to the data (the parameter $m$ is given in legend).}
    \label{fig:variances-linear}
\end{figure}
\begin{figure}[h]
    \centering
    \includegraphics[width=0.95\columnwidth]{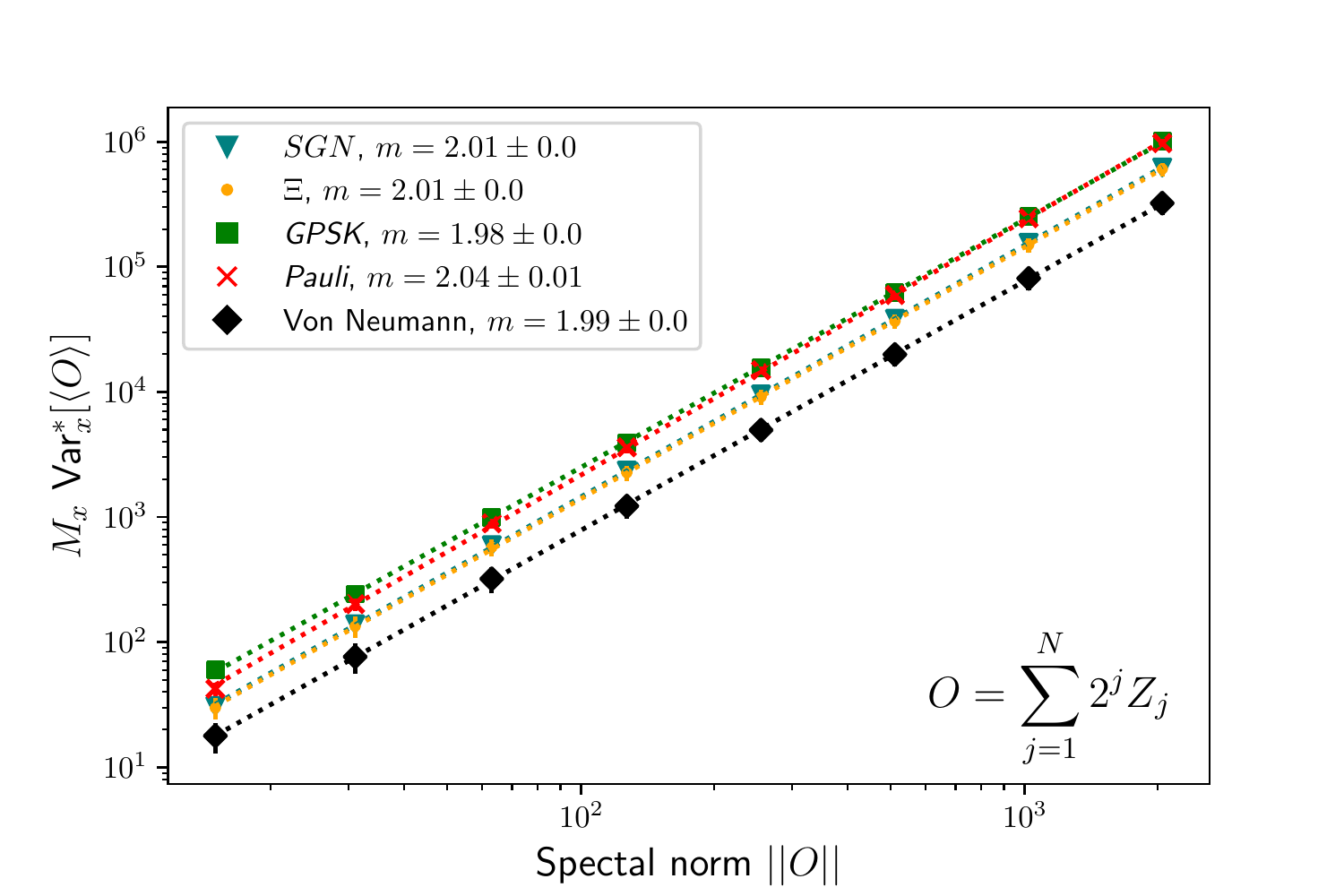}
   \caption{Comparison study of variances of different decompositions on random states generated by a hardware-efficient ansatz (see text for details). Different colours correspond to different decompositions [Eq.\eqref{eq:Odecomp}] of the target operator $O$ (see text for the description of all decompositions). Dashed lines are exponential-power-law fits ($\exp(aN^2+bN+c)$) to the data (the dominant scaling parameter $a$ is given in the legend).}
    \label{fig:variances-powers2}
\end{figure}

We measure the variances on random states generated by hardware-efficient ansatzes using PennyLane \cite{pennylane}. For each value of $N$, 100 random set of parameters (and therefore 100 random states) are generated and measured for all decompositions. For each decomposition $X$, we first use $10^5$ shots (allocated proportionally to the weight of each term) to obtain a rough estimate of the expectation value of each term $\expval{\Re(U_x)}$ for $x \in X$. These values are plugged in Eq.~\eqref{eq:shot-allocation} to get an estimate of the optimal shot allocation ratios $r_x = \frac{m_x}{M_x}$.
The variance of each term $\Var^*[\expval{\Re(U_x)}]$ is obtained by Eq.~\eqref{eq:Varstar} (or by sampling in the case of the QSP-approximation decomposition `SGN'). With these we compute the final shot-variance $M_X \Var^*_X[\expval{O}] = \sum_{x\in X} r_x^{-1} \Var^*[\expval{\Re(U_x)}$. 
Finally, we average the values of $M_X \Var^*_X[\expval{O}]$ obtained for each random state. This average is the quantity reported in Fig.~\ref{fig:variances-constant}, Fig.~\ref{fig:variances-linear} and  Fig.~\ref{fig:variances-powers2}.

The terms $\Xi_x$ are constructed as per Eq.~\eqref{eq:xi-decomp} using the known eigenvectors of $O$, and projectively measured on the prepared state (as these are reflection operators, Hadamard test samples match projective measurement samples).
The terms in the Pauli decomposition are also directly measured on the prepared state.
The GPSK-decomposition is constructed as described in \ref{app:GPSK} and measured through a Hadamard test.
The Von Neumann variance $\Var[O]$ is computed analytically.

The QSP approximation of $\Xi$ (denoted SGN from the sign term approximation) is implemented as described in Appendix~\ref{app:sign-decomposition} for $R=20$ and $\delta=0$.
For fair comparison with the other methods, echo verification is not used. The comparison between the $\Xi$ and SGN decomposition shows how the approximation increased the final variance. (The approximation also introduces a bias, see Appendix~\ref{app:sign-decomposition}.

All the simulations assume Hadamard-test-based measurement in an ideal circuit simulation: no circuit-level noise is considered and EV is not implemented.

We additionally report scaling results for the shot-variances of two other observables, $O = \sum_j j Z_j$ and $O = \sum_j 2^j Z_J$. The overall scaling of all decompositions matches the scaling of the operator norm $\lVert O \rVert$. Similarly to the case of Fig.~\ref{fig:variances-constant}, the $\Xi$ decomposition performs best, the SGN approximation has a relatively small effect on the shot-variance, and the Pauli decomposition shows the worst scaling.

\end{document}